\documentclass{article}
\usepackage{spconf,amsmath,graphicx,verbatim}
\usepackage{subcaption, comment, multirow, multicol, makecell, amssymb, bm, textcomp, kotex, arydshln}
\usepackage{adjustbox}
\usepackage{mathtools}
\usepackage{cite, url}
\usepackage{tabularx, makecell}
\usepackage{hyperref}
\usepackage{xcolor}
\usepackage{tabularx}
\hypersetup{
    colorlinks=true,
    linkcolor=purple,
    filecolor=magenta,      
    urlcolor=purple,
}
\newcolumntype{Y}{>{\centering\arraybackslash}X}
\usepackage[subtle]{savetrees}

\title{CONVOLUTION CHANNEL SEPARATION AND FREQUENCY SUB-BANDS AGGREGATION FOR MUSIC GENRE CLASSIFICATION}
\name{Jungwoo Heo$^*$\thanks{$^*$Equal contribution}, Hyun-seo Shin$^*$, Ju-ho Kim, Chan-yeong Lim, and Ha-Jin Yu\sthanks{$^\dag$Corresponding author.}\thanks{This research was supported by Basic Science Research Program through the National Research Foundation of Korea(NRF) funded by the Ministry of Science, ICT \& Future Planning (2020R1A2C1007081)}}
\address{School of Computer Science, University of Seoul}

\begin{document}

\ninept
\maketitle
\begin{abstract}
In music, short-term features such as pitch and tempo constitute long-term semantic features such as melody and narrative. 
A music genre classification (MGC) system should be able to analyze these features. 
In this research, we propose a novel framework that can extract and aggregate both short- and long-term features hierarchically. 
Our framework is based on ECAPA-TDNN, where all the layers that extract short-term features are affected by the layers that extract long-term features because of the back-propagation training. 
To prevent the distortion of short-term features, we devised the convolution channel separation technique that separates short-term features from long-term feature extraction paths. 
To extract more diverse features from our framework, we incorporated the frequency sub-bands aggregation method, which divides the input spectrogram along frequency bandwidths and processes each segment. 
We evaluated our framework using the Melon Playlist dataset which is a large-scale dataset containing 600 times more data than GTZAN which is a widely used dataset in MGC studies. 
As the result, our framework achieved 70.4\% accuracy, which was improved by 16.9\% compared to a conventional framework. 
\end{abstract}
%

\section{Introduction}
The significant amount of music in the digital music stores means that categorization is emerging as a crucial problem. 
Music genres, such as classical, jazz, pop, and rock, are conventional categories that cluster various music pieces. 
Many recent music genre classification (MGC) studies have confirmed that deep neural network (DNN)-based systems can achieve outstanding performance \cite{rajanna2015deep, bahuleyan2018music, ghildiyal2020music, pelchat2020neural}. 

Many previous studies have explored DNN structures suitable for the MGC tasks \cite{szegedy2015going}. 
Musical information is composed of short- and long-term information. 
Short-term information, such as pitch and tempo, are combined over time to form long-term information such as melody and narrative. \cite{liu2021bottom}. 
Therefore, the system that performs MGC should be able to extract short- and long-term information from music and understand their hierarchical relationship. 
From this perspective, Liu \textit{et. al.} proposed the bottom-up broadcast neural network (BBNN) framework that arranges convolution layers with different kernel sizes in parallel and aggregates output feature maps (similar to the inception \cite{he2016deep} block) \cite{liu2021bottom}. 
The BBNN showed superior performance at MGC, and from its results, we conjecture that the method that extracts short-term and long-term information in parallel would be practical for MGC. 

This paper describes our goal to devise a effective framework for the MGC task. 
To this end, we use ECAPA-TDNN \cite{desplanques2020ecapa} as the baseline. 
ECAPA-TDNN, designed for speaker verification, has achieved state-of-the-art performance by extracting and incorporating features in various time scales. 
Our reasons for employing ECAPA-TDNN are as follows: \textbf{i)} it consists of SE-Res2Block \cite{gao2019res2net} with various dilations that extract short- and long-term features \textbf{ii)} it concatenates and condenses the outputs of all SE-Res2Blocks hierarchically. 
In addition, we propose a convolution channel separation (CCS) technique to transform ECAPA-TDNN to be suitable for MGC tasks. 
During the training phase, the DNN optimizes the weights through the backpropagation. 
Since the back-propagation goes from the upper to the lower layer, the upper layers can affect the lower layers. 
In ECAPA-TDNN, the output of the lower SE-Res2Block is fed to the upper SE-Res2Block. 
However, low locality features (short-term features) are crucial to the MGC task \cite {choi2017transfer}. 
Therefore, we assumed that retaining information from several lower layers would be advantageous for performing the MGC task. 
The proposed CCS divides the convolution channels into ``continuous" and ``stop" processing channels. 
Only the continuous-processing channels are fed to the next SE-Res2Block, and stop-processing channels are directly passed to the pooling layer. 
Thus, stop-processing channels keep information from lower layers without being affected by higher layers.

To improve our framework further, we also incorporated the frequency sub-bands aggregation (FSA) method. 
The FSA slices a spectrogram into sub-bands along the frequency axis and processes each sub-band. 
Studies in various fields (spoofing detection, speech enhancement, and acoustic scene classification) demonstrated that the FSA technique could improve the performance \cite{chettri2020subband, yu2022dmf, phaye2019subspectralnet}. 
Since music is composed of several instruments in different frequency bands, we expect to extract more diverse sepctral information by applying the FSA technique. 
Therefore, we modified and applied the FSA method to enhance the accuracy of our framework. 

Many MGC studies tested their systems with a small data sets because it can be difficult to obtain data due to music copyrights. 
For this study, we evaluated our proposed framework using a new large-scale dataset called the Melon Playlist dataset \cite{ferraro2021melon}. 
The GTZAN \cite{tzanetakis2002musical} data set is widely used in MGC studies, and this data set contains 100 samples in total for evaluation. 
Meanwhile, the Melon Playlist dataset contains approximately 600 times more samples than the GTZAN dataset. 

Through this study, we made the following contributions. 
\begin{itemize}
    \item We grafted ECAPA-TDNN into MGC studies. 
    \item We designed the CCS method to prevent high-level features from interfering with low-level features in ECAPA-TDNN. 
    \item We applied FSA to extract more diverse information from different frequency bands. 
\end{itemize}

\section{ECAPA-TDNN}
\label{section:2}
In this section, we present the structure of ECAPA-TDNN and its architectural advantages in MGC task. 
Figure \ref{figure:ecapa} shows the overall architecture of ECAPA-TDNN, and Table \ref{table:ECAPA} describes the components of each block. 
ECAPA-TDNN converts the input spectrogram into a one-dimensional feature map through a convolution layer (Fst-Conv). 
And then, the converted feature map is fed to three SE-Res2Blocks with three different dilations. 
Finally, features extracted from these blocks are refined and aggregated through a convolution (Last-Conv) and an attentive statistics pooling (Pooling) layer. 

Musical information consists of short-term (such as pitch and tempo) and long-term (such as melody and narrative) information hierarchically \cite{choi2017transfer}. 
Therefore, a framework that can hierarchically extract and aggregate musical information might be advantageous for MGC \cite{liu2021bottom}. 
From this perspective, we conjecture that the ECAPA-TDNN \cite{desplanques2020ecapa} has the potential to handle musical information because of two components: i) SE-Res2Blocks with various dilations ii) Last-Conv and Pooling layers that used for feature aggregation. 
As shown in Figure \ref{figure:ecapa}, ECAPA-TDNN contains the SE-Res2Blocks B1, B2, and B3 with 2, 3, and 4 dilation spacing, respectively. 
In addition, each SE-Res2Block receives the sum of the outputs of all previous blocks that may allow each layer to exploit features from previous blocks. 
Dilated convolution with different dilation spacings are beneficial for the network to capture time scale contexts \cite{yu2015multi}. 
Due to this characteristics, SE-Res2Blocks can be advantageous in extracting multiple time scale information and exploiting various term information extracted from the previous layers. 
In other words, this structure allows the ECAPA-TDNN to extract short- and long-term information. 
Meanwhile, the ECAPA-TDNN has a process that concatenates and aggregates the output of multiple layers. 
As depicted in Figure \ref{figure:ecapa}, the outputs of B1, B2, and B3 are concatenated and input to the Last-Conv. 
Then, the Last-Conv refines the feature map for Pooling to aggregate appropriate information required for genre classification. 
Through this process, ECAPA-TDNN hierarchically analyzes short- and long-term features. 

\section{Proposed Method}
The purpose of this research is to develop a suitable framework for the MGC. 
We propose and apply the convolution channel separation and frequency sub-bands aggregation methods for MGC task, beyond simply using the ECAPA-TDNN as a foundation. 
In the following subsections, we describe the motivation for using these techniques and each technique's process in detail. 

\subsection{Convolution channel separation}
\begin{figure}[!ht]
\begin{center}
    \centering
    \includegraphics[width=\linewidth]{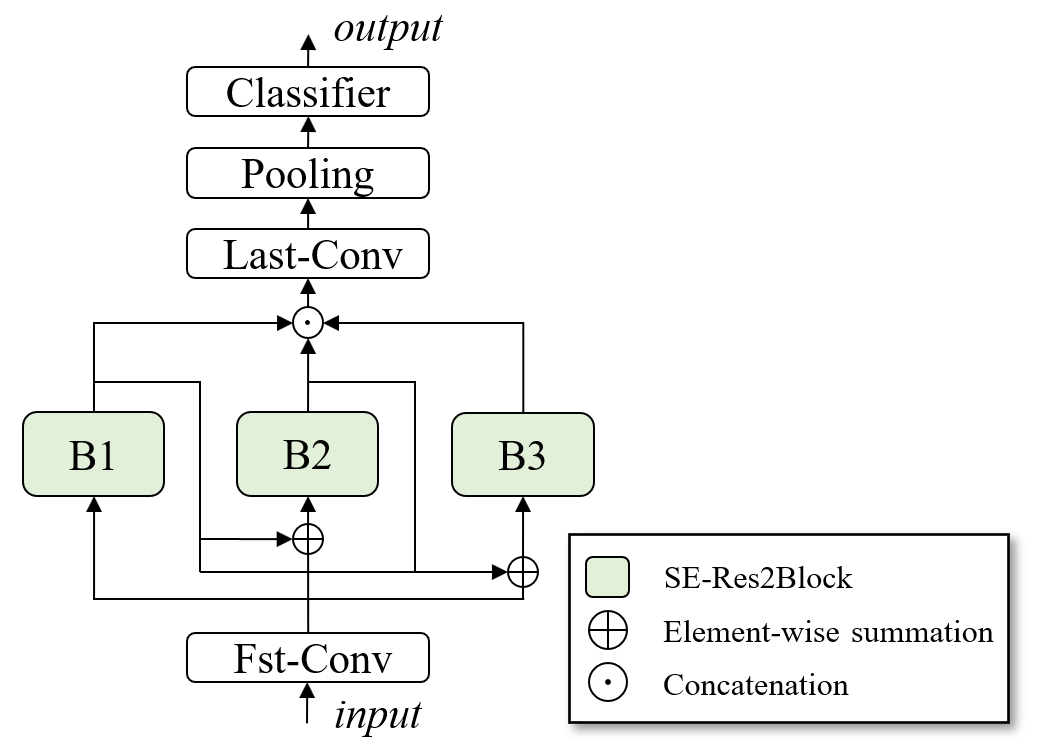}
    \vspace{-0.2cm}
    \caption{Overall architecture of ECAPA-TDNN. B1, B2, and B3 denote SE-Res2Blocks.}
\label{figure:ecapa}
\end{center}
\end{figure}

In general, DNN optimizes their weights through back-propagation. 
The back-propagation process proceeds from the upper layer to the lower layer. 
As shown in Figure \ref{figure:ecapa} the output of B1 is input to B2 and B3, and the output of B2 is input to B3. 
That means B1 and B2 is a lower layer than B2 and B3, respectively, and they are affected by higher layers. 
According to Choi \textit{et al.}  \cite{choi2017transfer}, features extracted from low-level layers are required to perform the MGC task. 
Considering the findings, we conjecture preserving some low-level information would be advantageous for identifying music genres. 
Therefore, we designed and applied the CCS method to ECAPA-TDNN.

\begin{table}[!ht]
\caption{
Components of each block in ECAPA-TDNN. 
Res2 Conv1D(k, s, d) is the Res2Net\cite{gao2019res2net} one-dimensional convolution layer with kernel size k, stride size s, and dilation spacing d. 
ASP and SE-Block denote attentive statistics pooling and Squeeze-and-Excitation block. ReLU and batch normalization layers exist after every Conv and Res2 Conv layer. 
}
\centering
\label{table:ECAPA}
\resizebox{\linewidth}{!}{
\begin{tabular}{ccc}
\Xhline{2\arrayrulewidth}
\textbf{Layer} & \textbf{Block structure} & \textbf{Output} \\
\Xhline{2\arrayrulewidth}
Fst-Conv & Conv1D(5, 1, 1) & 1024 $\times$ T \\
\hline
\multirow{4}{*}{SE-Res2Block (B1)}& Conv1D(1, 1, 1) &  128 $\times$ T\\
& Res2 Conv1D(3, 1, 2) & 128 $\times$ T \\
& Conv1D(1, 1, 1) &  1024 $\times$ T \\
& SE-Block &  1024 $\times$ T \\
\hline
\multirow{4}{*}{SE-Res2Block (B2)}& Conv1D(1, 1, 1) &  128 $\times$ T \\
& Res2 Conv1D(3, 1, 3) & 128 $\times$ T \\
& Conv1D(1, 1, 1) &  1024 $\times$ T \\
& SE-Block &  1024 $\times$ T \\
\hline
\multirow{4}{*}{SE-Res2Block (B3)}& Conv1D(1, 1, 1) & 128 $\times$  T\\
& Res2 Conv1D(3, 1, 4) & 128 $\times$ T \\
& Conv1D(1, 1, 1) &  1024 $\times$  T\\
& SE-Block &  1024 $\times$  T\\
\hline
Last-Conv & Conv1D(1, 1, 1) & 1536 $\times$  T \\
\hline
Pooling & ASP & 3072 $\times$ 1\\
\hline
Classifier & FC & Num of classes \\
\Xhline{2\arrayrulewidth}
\end{tabular}
}
\end{table}

\begin{figure}[t]
\begin{center}
    \centering
    \includegraphics[width=\linewidth]{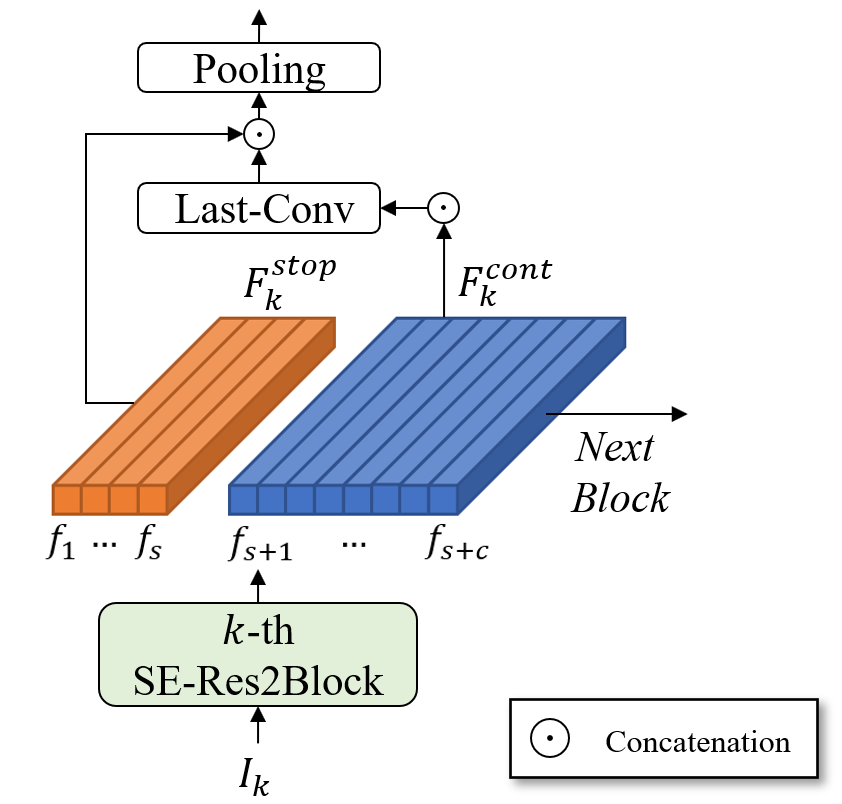}
    \vspace{-0.2cm}
    \caption{Structure of $k$-th SE-Res2Block with the proposed convolution channel separation method applied to the ECAPA-TDNN. ($F^{stop}$: the set of stop-processing channels, $F^{cont}$: the set of continuous-processing channels, $f_i$: channel element of set $F^{stop}$ or $F^{cont}$, which is a vector of the feature map, $I$: input.)}
\label{figure:ccs}
\end{center}
\vspace{-2em}
\end{figure}

Figure \ref{figure:ccs} illustrates the structure of \textit{k}-th SE-Res2Block with the proposed CCS method, and Equations (1)-(3) explain the process of CCS. 
The input spectrogram is converted to a feature map $X$ through the Fst-Conv layer. 
Equation (1) denotes the input of the \textit{k}-th SE-Res2Block. 
\begin{gather}
    I_{k} =
    \begin{cases}
    X + \Sigma_{i=1}^{k-1}F_i^{cont} ,& k > 1 \\
    X ,& k = 1
    \end{cases}
\end{gather}
The \textit{k}-th SE-Res2Block receives $I_k$ and outputs $F_k$ ($k \in \{1, 2, 3 \}$), which is the set of the feature vector $f$. 
Then, as depicted in Figure \ref{figure:ccs}, $F_k$ is divided into two subsets $F^{stop}_k$ and $F^{cont}_k$, which contain $s$ and $c$ elements, respectively. 
We define $F^{stop}_k$ and $F^{cont}_k$ in tabular form as in Equations (2) and (3). 
\begin{gather}
    F_k^{stop} = \{ f_1, \dots, f_{s}\} \\ 
    F_k^{cont} = \{ f_{s+1}, \dots, f_{s + c}\}
\end{gather}
After splitting the channel, we directly deliver the features of stop-processing channels ($F^{stop}_k$) to the Pooling layer. 
On the other hand, the features of continuous-processing channels ($F^{cont}_k$) are fed to the next SE-Res2Block and Last-Conv. 
As an exception, the last features of continuous-processing channels are fed only to Last-Conv. 
In the feature aggregation process, all continuous-processing channels ($F^{cont}_1$, $F^{cont}_2$, $F^{cont}_3$) are concatenated and refined through Last-Conv. 
Then, the Pooling layer condense all stop-processing channels ($F^{stop}_1$, $F^{stop}_2$, $F^{stop}_3$) and the output of Last-Conv.

In the CCS technique, the output feature map ($F$) is divided into two parts ($F^{cont}$ and $F^{stop}$) and an additional processing is applied to one of the parts ($F^{cont}$). 
Therefore, this method encourages the framework to decide which information to pass on to the upper layer and which information to keep.

\subsection{Frequency sub-bands aggregation}
\begin{figure}[t]
\begin{center}
    \centering
    \includegraphics[width=\linewidth]{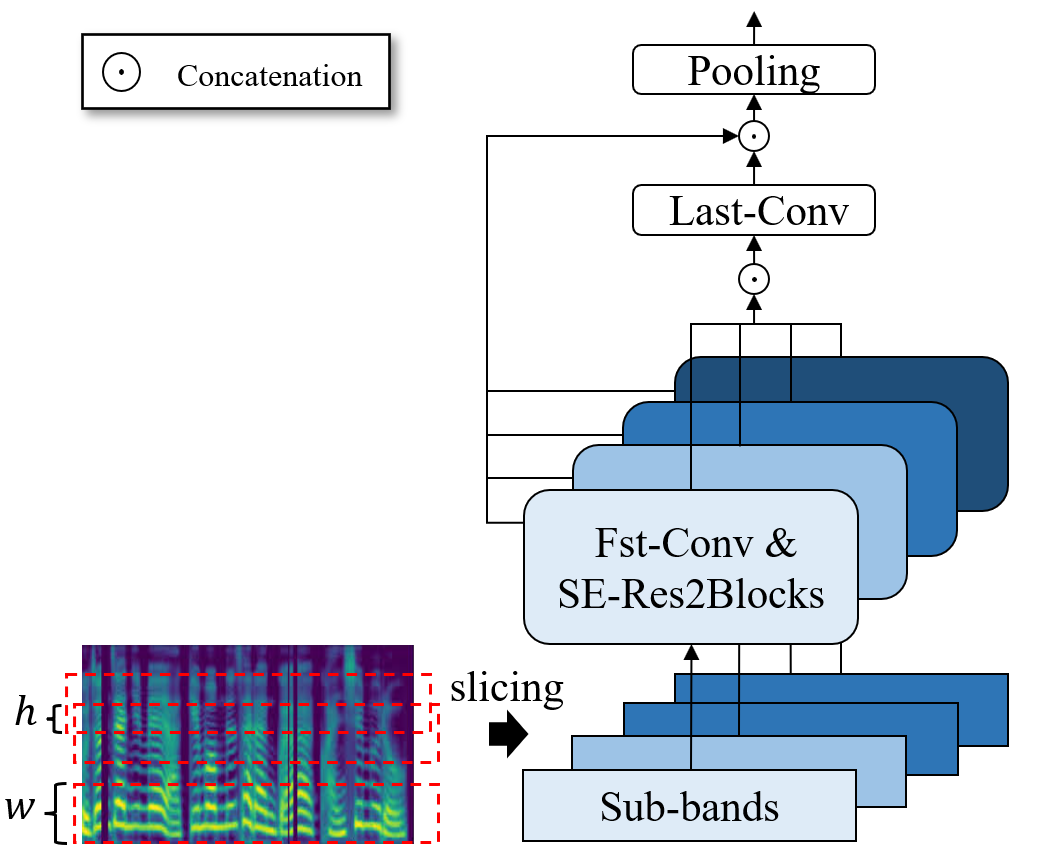}
    \vspace{-0.2cm}
    \caption{Illustration of the frequency sub-bands aggregation method applied to our framework. ($w$: window size, $h$: hop size.)}
\label{figure:fsa}
\end{center}
\vspace{-2em}
\end{figure}

The FSA approach divides a spectrogram into sub-bands and processes each sub-band. 
This method's effectiveness has been demonstrated in various tasks such as spoofing detection, speech enhancement, and acoustic scene classification \cite{rajanna2015deep, bahuleyan2018music, ghildiyal2020music, pelchat2020neural}. 
We consider that FSA method would also be suitable for MGC tasks because music consists of several instruments with different frequency bands. 
Therefore, we applied this technique to our framework, as depicted in Figure \ref{figure:fsa}. 

The operation process of the applied FSA is as follows. 
First, we divided the input spectrogram into four overlapping segments as shown at the bottom of Figure \ref{figure:fsa}. 
When slicing the spectrogram, we set the window size ($w$) to 18 and the hop size ($h$) to 10. 
We also duplicated the feature extraction module (consisting  of Fst-Conv and three SE-Res2Blocks) to four pieces. 
The segments were fed to the four different Fst-Conv individually, and they are processed into short- and long- term features. 
The output feature maps are input to B1, B2, and B3, connected to each Fst-Conv. 
After that, the continuous-processing channels extracted from the four different modules are concatenated and fed to the Last-Conv and the stop-processing channels are transmitted directly to the Pooling layer. 
Both the Last-Conv and Pooling layers aggregate features extracted from different frequency bands. 
Applying the FSA technique, we reduced the channels of Fst-Conv, B1, B2, and B3 to adjust the number of parameters to the same level as the original ECAPA-TDNN.

\section{Experiment}
\subsection{Dataset}
We used two datasets in our research. 
The first is the GTZAN dataset. 
The GTZAN dataset was published in 2002 and has been widely used in MGC research; it consists of 1,000 songs evenly distributed into ten different genres. 
Each sample is 30 s in length and the sample rate is 22,050 Hz at 16 bit. 
We partitioned the training and test sets following the k-fold process. 
The second is the Melon Playlist dataset \cite{ferraro2021melon}. 
The Melon Playlist dataset was released in 2021 to promote the study of music playlist prediction. 
It was not designed for the MGC task, but we used as such. 
The dataset consists of 649,091 songs and each song's metadata (album, artist, genre, etc.). 
Genres are divided into 30 general labels (e.g., jazz, pop, and classical) and 224 detailed labels (e.g., 1980s and 1990s); however, we only used the general genre label for the study. 
Each genre includes a different number of songs. 
Each of the 30 genres is marked as ``GN\_\_00" (for example, ``GN0900" denotes pop and ``GN1700" denotes jazz). 
Note that we fixed ``GN9000" as ``GN3000" because ``GN3000" was incorrectly marked as ``GN9000". 
In addition, some samples have either none or multi genre labels. 
Therefore, in the process of removing this ambiguous data, we excluded approximately 50,000 samples and two genre labels, ``GN1500" and  ``GN2500", which were ``K-pop idol group" and ``OST," respectively.  
All samples were of 20$\sim$50s length, and were pre-processed to the Mel-spectrogram with the following settings: 16 kHz sampling rate, window and hop size of 512 and 256 samples, Hann window function, and 48 filter bands. 
We divided the training and test sets following the k-fold manner, as for GTZAN. 
The codes can be found on our Github \footnote[1]{https://github.com/Jungwoo4021/ECAPAwithCCSandFSA}.

\begin{table}[!ht]
\caption{
Comparison of the classification accuracy (\%) between the proposed frameworks and the recently proposed models on the GTZAN dataset. 
The CCS refers the convolution channel separation method with stop-processing channel size of $s$ and continuous-channel size of $c$. 
The FSA and TTA denote the frequency sub-bands aggregation approach and test time augmentation method\cite{chung2018voxceleb2}, respectively. 
}
\centering
\resizebox{\linewidth}{!}{%
\label{table:gtzan}
\begin{tabular}{l|c}
\Xhline{2\arrayrulewidth}
\textbf{Models} & \textbf{Accuracy (\%)} \\
\hline
\#1 AuDeep (Freitag et al.\cite{freitag2017audeep}) & 85.4 \\
\#2 NNet2 (Zhang et al.\cite{zhang2016improved}) & 87.4 \\
\#3 CVAF (Nanni et al.\cite{nanni2017combining}) & 90.9 \\
\#4 BBNN (Liu et al.\cite{liu2021bottom}) & 93.9 \\
\hline
\#5 ResNet18 & 85.7 \\
\#6 ResNet34 & 86.3 \\
\#7 SE-ResNet34 & 85.3 \\
\#8 ECAPA-TDNN & 87.9 \\
\#9 ECAPA-TDNN + CCS ($s$=512, $c$=512) & 87.9 \\
\#10 ECAPA-TDNN + CCS ($s$=256, $c$=768) & 88.2 \\
\#11 ECAPA-TDNN + CCS ($s$=256, $c$=1024) & 88.6 \\
\#12 ECAPA-TDNN + CCS ($s$=512, $c$=1024) & 88.6 \\
\#13 ECAPA-TDNN + CCS ($s$=256, $c$=768), FSA & 89.5 \\
\#14 ECAPA-TDNN + CCS ($s$=256, $c$=1024), FSA & 89.7 \\
\#15 ECAPA-TDNN + CCS ($s$=512, $c$=1024), FSA & 90.2 \\
\#16 ECAPA-TDNN + CCS ($s$=512, $c$=1024), FSA (TTA) & 91.1 \\
\Xhline{2\arrayrulewidth}
\end{tabular}}
\end{table}

\subsection{Experiment setting}
We used 202 frames of 48-dimensional Mel-spectrogram as an input. 
All models were trained with an Adam optimizer and learning rate $10^{-3}$.  

The learning rate was reduced to $10^{-6}$ for 80 epochs by a cosine annealing \cite{loshchilov2016sgdr}. 
We used categorical cross-entropy loss and set the batch sizes to 256 and 64 in Melon and GTZAN, respectively. 

The primary metric of MGC research is accuracy. 
To this end, all experiments were performed with 10-fold cross-validation following previous studies' evaluation protocols  \cite{liu2021bottom, choi2017transfer}. 
Experiments were implemented based on the PyTorch framework \cite{paszke2019pytorch}. 

\subsection{Results}

We explored various frameworks in MGC tasks with the GTZAN and Melon Playlist datasets. 
\\\\
\textbf{GTZAN dataset.} \; Table \ref{table:gtzan} shows the accuracy for the GTZAN dataset. 
Experiments \#1-4 show previous studies' results and \#5-16 show the performance of our frameworks. 
To compare the performance of various frameworks on the GTZAN dataset, we evaluated four frameworks (\#5-8). 
In experiments on \#5-7, simply adopting a conventional framework for the MGC task was not effective. 
The vanilla ECAPA-TDNN shown in \#8 revealed an accuracy of 87.9\%, which outperforms conventional frameworks, such as ResNet18, ResNet34, and SE-ResNet34. 
Experiments \#9-12 demonstrate the results of ECAPA-TDNN grafted with the proposed CCS. 
As exhibited in experiments with the four settings by changing the size of stop-processing channel $s$ and continuous-channel $c$, all except \#9 showed improved performance compared to vanilla ECAPA-TDNN.  
In \#13-16 grafted with FSA, the performance was further enhanced, achieving 89.5$\sim$91.1\%. 
These results suggest that our model has competitive performance. 
\\\\
\textbf{Melon Playlist dataset.} \; In general, more training data provides improved generalization performance, and more evaluation data provide more reliable performance. 
The GTZAN dataset has been widely used in MGC research, but due to its small scale, the evaluation data may not be sufficient for comparing the framework's generalization performance. 
Therefore, we measured the frameworks' performance on the Melon Playlist dataset, which is 600 times larger.

\begin{table}[t]
\caption{
Classification accuracy (\%) on the Melon Playlist dataset is compared across various frameworks. 
}
\centering
\resizebox{\linewidth}{!}{%
\label{table:melon}
\begin{tabular}{l|c}
\Xhline{2\arrayrulewidth}
\textbf{Models} & \textbf{Accuracy (\%)} \\
\hline
\#17 VGGNet & 45.3 \\
\#18 ResNet18 & 63.6 \\
\#19 ResNet34 & 63.6 \\
\#20 SE-ResNet34 & 64.1 \\
\#21 BBNN* & 60.2 \\
\#22 ECAPA-TDNN & 68.8 \\
\hline
\#23 ECAPA-TDNN + CCS ($s$=256, $c$=1024) & 69.2 \\
\#24 ECAPA-TDNN + CCS ($s$=256, $c$=768) & 69.6 \\
\#25 ECAPA-TDNN + CCS ($s$=512, $c$=512) & 69.5 \\
\#26 ECAPA-TDNN + CCS ($s$=256, $c$=1024), FSA & 70.3 \\
\#27 ECAPA-TDNN + CCS ($s$=256, $c$=768), FSA & \textbf{70.4} \\
\Xhline{2\arrayrulewidth}
\end{tabular}}
\end{table}

Table \ref{table:melon} delivers the accuracy of the frameworks in the Melon Playlist evaluation protocol. 
We implemented and explored several frameworks (\#17-22) for the Melon Playlist dataset. 
Predictably, traditional frameworks were not effective on this dataset either(\#17-20). 
It is noteworthy that the BBNN models (\#5) shows 60.2\% accuracy.  
This result contrasts with the superior performance of 93.9\% in the GTZAN dataset, indicating that severe performance degradation occurs in the BBNN. 
We analyze that the BBNN suffers such a performance degradation because the number of parameters is insufficient for the learning capacity of a large dataset (the total number of parameters is about 180,000). 
Meanwhile, in \#22, the ECAPA-TDNN achieved the best performance among the frameworks. 
Through these results, we confirm that ECAPA-TDNN is an effective framework for MGC task regardless of dataset. 

Subsequently, we conducted experiments to verify the effectiveness of the proposed CCS and FSA methods. 
When applying the CCS technique, as in \#23-25, these frameworks displayed improved performances compared to vanilla ECAPA-TDNN, as with the previous tendency. 
Through this, we confirm that the CCS technique is effective for MGC. 
Experiments \#26 and \#27 show the experimental results when FSA is grafted with CCS. 
Similarly, both experiments exhibit better performance than that with only CCS; finally, the highest accuracy of 70.4\% was achieved using our proposed framework. 

We analyzed these results as follows.
The proposed framework has a valid learning capacity on large datasets and is suitable for MGC. 
CCS and FSA technologies designed in consideration of music's characteristics can contribute to improving the MGC performance. 

\section{Conclusion}
We proposes a novel framework for the MGC task. 
We conjectured that the structural characteristics of ECAPA-TDNN would be advantageous for performing MGC; on this basis, we devised a novel framework by applying CCS and FSA techniques. 
The proposed CCS divides a feature map into two parts, and one part is processed more to extract long-term information, while the other is fed directly to the upper layer to preserve short-term information. 
As we conjectured, ECAPA-TDNN showed the best accuracy compared to other frameworks using the large Melon Playlist dataset. 
Furthermore, the proposed framework achieved the highest accuracy for the same dataset. 
The original ECAPA-TDNN showed 68.8\% accuracy, while the proposed model showed 70.4\%. 
This result demonstrates that the proposed techniques can extract music information more effectively for classifying music genres. 
In addition, we also evaluated several conventional frameworks in the Melon Playlist dataset. 
As the future works, we plan to analyze the factors for the performance improvement experimentally and achieve higher accuracy by exploring more suitable frameworks. 

\bibliographystyle{IEEEbib}
\bibliography{refs}

\end{document}